\documentclass[letter]{aa} 

\usepackage{graphicx}
\usepackage{txfonts}

\usepackage{hyperref}
\usepackage[inline]{enumitem} 
\usepackage{multirow}

\begin{document}

   \title{The puzzle of isolated and quenched dwarf galaxies in cosmic voids}
   \titlerunning{Isolated and quenched dwarf galaxies in voids}


   \author{Bahar Bidaran\inst{\ref{ugr1}}
   \and Isabel Pérez\inst{\ref{ugr1},\ref{ugr2}}
   \and Laura Sánchez-Menguiano\inst{\ref{ugr1},\ref{ugr2}}
   \and María Argudo-Fernández\inst{\ref{ugr1},\ref{ugr2}}
    \and Anna Ferr\'e-Mateu\inst{\ref{iac},\ref{ull}}
    \and Julio F. Navarro \inst{\ref{UBC}}
    \and Reynier F. Peletier\inst{\ref{kapteyn}}
    \and Tomás Ruiz-Lara\inst{\ref{ugr1},\ref{ugr2}}
     \and Glenn van de Ven\inst{\ref{Vienna}} 
     \and Simon Verley\inst{\ref{ugr1},\ref{ugr2}}
     \and Almudena Zurita\inst{\ref{ugr1},\ref{ugr2}}  
     \and Salvador Duarte Puertas\inst{\ref{ugr1},\ref{ugr2},\ref{laval}}
     \and Jes\'us Falc\'on-Barroso\inst{\ref{iac},\ref{ull}}
      \and Patricia S\'anchez-Bl\'azquez\inst{\ref{ucm}}
       \and Andoni Jiménez\inst{\ref{ugr1}}}

   \institute{
   Dpto. de F\'{\i}sica Te\'orica y del Cosmos, Facultad de Ciencias (Edificio Mecenas), University of Granada, E-18071, Granada, Spain\label{ugr1}
    \and Instituto Carlos I de F\'\i sica Te\'orica y Computacional, Universidad de Granada, E18071, Granada, Spain\label{ugr2}
    \and Instituto de Astrof\'isica de Canarias, c/V\'ia L\'actea s/n, E-38205, La Laguna, Tenerife, Spain\label{iac}
\and Departamento de Astrof\'isica, Universidad de La Laguna, E-38206, La Laguna, Tenerife, Spain\label{ull}
    \and Department of Physics and Astronomy, University of Victoria, Victoria, BC V8P 5C2, Canada \label{UBC} 
    \and Kapteyn Astronomical Institute, University of Groningen, PO Box 800, 9700 AV Groningen, The Netherlands\label{kapteyn}
    \and Department of Astrophysics, University of Vienna, T\"urkenschanzstrasse 17, 1180 Vienna, Austria \label{Vienna}
    \and D\'epartement de Physique, de G\'enie Physique et d’Optique, Universit\'e Laval, and Centre de Recherche en Astrophysique du Qu\'ebec (CRAQ), Québec, QC, G1V 0A6, Canada\label{laval}
    \and Departamento de Física de la Tierra y Astrofísica, Universidad Complutense de Madrid, E-28040 Madrid, Spain\label{ucm}}
    
   \date{Received Month Day, Year; accepted Month Day, Year}

  \abstract
   {We report, for the first time, the detection of a sample of quenched and isolated dwarf galaxies (with 8.9 $<$ log(M$_{\rm \star}$/M$_{\rm \odot}$) $<$ 9.5) in the least dense regions of the cosmic web, including voids, filaments, and walls. These dwarfs have no neighbouring galaxy within 1.0~Mpc in projected distance. Based on the full spectral fitting of their central spectra using Sloan Digital Sky Survey data, these galaxies are gas-deprived, exhibit stellar mass assembly very similar to dwarfs in the central regions of galaxy clusters, and have experienced no significant star formation in the past 2 Gyr. Additionally, analysis of r-band images from the Dark Energy Camera Legacy Survey showed that these dwarf galaxies host a central Nuclear Star Cluster (NSC). Detecting quenched, isolated dwarf galaxies in cosmic voids indicates that environmental factors are not the sole drivers of their quenching. Internal mechanisms, such as feedback from in-situ star formation, also contributing to the NSC formation, black holes, or variations in conditions during their formation, offer potential explanations for star formation suppression in these galaxies. These findings highlight the need for a significant revision in our understanding of baryonic physics, particularly concerning the formation and evolution of low-mass galaxies. }

   \keywords{Galaxies: dwarf -- Galaxies: evolution -- Galaxies: star formation -- large-scale structure of Universe}

   \maketitle
%

\section{Introduction}
Dwarf galaxies, defined as galaxies with log(M$_{\rm \star}$/M$_{\rm \odot}$) $<$ 9.5 \citep[e.g.,][]{2018Eigenthaler}, represent the most abundant galaxy type in the Universe. The observed morphological segregation highlights a significant increase in the number density of quenched dwarf galaxies with early-type morphologies within high-density environments, such as galaxy clusters, compared to lower-density regions of the cosmic web, including voids, walls, and filaments \citep[e.g.,][]{1980Dressler,1995Caon,2017Ricciardelli}, emphasising the sensitivity of these galaxies to environmental effects. In high-density environments, ram pressure stripping can rapidly deplete dwarf galaxies of their cold gas and stop star formation within about 1 Gyr, while tidal interactions over several Gyr can further influence their star formation history and morphology \citep{2006Boselli, 2014Boselli, 2022Boselli}. It is widely accepted that environmental factors are the primary drivers of star formation suppression in dwarf galaxies \citep[e.g.,][]{2014Wetzel} and quenched dwarf galaxies, particularly with log(M$_{\rm \star}$/M$_{\rm \odot}$) $<$ 9.0, do not exist in isolation \citep[e.g.,][]{2012Geha}. 

However, attributing quenching solely to the present-day environment is an oversimplification, as it fails to account for the existence of gas-deprived dwarf galaxies with old stellar populations and early-type morphologies located beyond the clusters virial radius \citep[e.g.,][]{2002Lewis,2012Smith,2015Haines}. Galaxies can be accreted into clusters not only individually but also as members of smaller groups or along filaments. Therefore, recent studies have proposed the idea of pre-processing under environmental effects in pre-cluster environments to address these observations \citep[e.g.,][]{2008vandenBosch,2021Winkel, 2021Bidaran}. Recent observations and cosmological hydrodynamical simulations have highlighted the potential role of massive black holes (BHs) in suppressing star formation in dwarf galaxies \citep[e.g.,][]{2018Penny, 2023Sharma}. Meanwhile, theoretical models suggest that supernova (SNe) feedback has a relatively limited impact on quenching dwarf galaxies with log(M${\rm_\star}$/M${\rm_\odot}$) $<$ 8.0 \citep{2008Valcke,2016Emerick}.

If non-environmental mechanisms contribute to the quenching of dwarf galaxies, we should be able to detect their signatures in quenched and isolated dwarf galaxies. Attempts to identify such systems, although different in quenching or isolation criteria, have predominantly yielded a near absence of these galaxies \citep{2012Geha,2017Janz}, and some resulted in discovering different types of galaxies, such as compact ellipticals \citep[e.g.,][]{2015Chilingarian,2017Janz}. Nonetheless, no systematic searches for quenched dwarf galaxies have yet been conducted in the least dense regions of the cosmic web, such as voids \citep[e.g.,][]{2001Peebles,2011Kreckel}. Voids are pristine environments \citep{2016Annibali} that encompass large volumes, averaging approximately 35 h$^{-1}$ Mpc in size \citep{2011Weygaert}. Existing studies, albeit limited in scope and focused on a small number of dwarf galaxies within voids, have consistently shown that all observed galaxies are actively forming stars \citep[e.g.,][]{2016Beygu,2023delosReyes}. 

In this letter, we present, for the first time, the detection of a sample of isolated and quenched dwarf galaxies located in cosmic voids, filaments, and walls, where they are presumably insulated from the environmental effects that dominate high-density regions. Section \ref{data} outlines the criteria used to identify these isolated and quenched dwarf galaxies, while Section \ref{analysis} details the spectral data analysis conducted. In Section \ref{results}, we discuss the characteristics of their star formation histories and associated properties, and in Section \ref{discussion}, we explore the potential mechanisms behind their quenching.

\section{Sample and data}\label{data}
We constructed the main sample of this work by mining two large galaxy catalogues which share the same redshift range and are based on the 7th data release of the Sloan Digital Sky Survey (SDSS-DR7), with magnitude completeness limit, set at r-Petrosian $<$ 17.77 mag \citep{2002Strauss,2015A&A...578A.110A}:

\rm\textbf{Void catalogue:} We used a subset sample of 8690 galaxies situated within 42 voids, exctracted from the catalogue of 1055 nearby voids presented in \cite{2012Pan}. This sample is a refined version of the Pan catalogue following similar criteria as the CAVITY project \citep[Calar Alto Void Integral-field Treasury surveY,][]{2024Perez}. This sample includes only those galaxies whose hosting voids fall entirely within the SDSS footprint and are confined to the redshift range of 0.005$<$ z $<$0.05.

\textbf{Filaments \& walls catalogue:} We used the sample of 15000 galaxies in filaments and walls of  \cite{2023Natur.619..269D} to identify dwarf galaxies inhabiting these cosmic structures. Given the comparable galaxy number density observed in cosmic walls and filaments \citep[e.g.,][]{2014Cautun}, for the purpose of our study, we treat them interchangeably as similar large-scale environments, referred to as filaments henceforth. 

\vspace*{2px}
From these extensive datasets, we specifically targeted dwarf galaxies with stellar masses log(M$_{\rm \star}$/M$_{\rm \odot}$) $<$ 9.5. {The stellar masses are provided by the MPA-JHU catalogue (complete in the redshift range of 0.005$<z<$0.22) and are estimated based on photometry, as reported in \citet{2004Tremonti,2004Brinchmann}}. The mass threshold leaves us with 4208 and 6473 dwarf galaxies in voids and filaments, respectively. It should be noted that while our stellar mass criterion encompasses lower-mass dwarf galaxies, both of the samples used here suffer from the SDSS magnitude completeness limit, corresponding to log(M$_{\rm \star}$/M$_{\rm \odot}$) $<$ 8.9. Hence, our analysis provides a lower limit for the isolated and quenched dwarf galaxies population and the final sample does not represent the actual size of quenched and isolated dwarf galaxy populations in the local Universe.

Following the red-sequence galaxy criterion outlined by \cite{2020Sales}\footnote{$g-r = 0.61 + 0.052(log_{10} (\frac{M_{\star}}{M_{\odot}})-10)$)}, we identified 503 red-sequence dwarf galaxies within voids and 768 within filaments. From these candidates, we selected dwarf galaxies that are exceptionally isolated. We investigated their local environment, using the NASA-Sloan Atlas (NSA\footnote{\url{https://www.sdss4.org/dr17/manga/manga-target-selection/nsa/}}) and following \cite{2015A&A...578A.110A} selection criteria in a velocity difference-projected distance space. Specifically, we selected those red-sequence dwarf galaxies for which no neighbour galaxy (brighter than  M$_{\rm r}$ $\sim$ $-$17 mag) was found within a projected distance of 1.0 Mpc and a line-of-sight velocity interval of $\Delta$V = 500 km/s.  Also, nearby objects (z$<$\,0.01) for which isolation estimates suffer from large uncertainties are discarded \citep{2007Verley}. These criteria are commonly used for defining samples of “isolated” or “field” galaxies in various studies \citep[e.g.,][]{2012Geha, 2021Dickey}. Only 157 galaxies in voids and 41 galaxies in filaments fulfil the isolation criteria.

\begin{figure*}
\centering
\includegraphics[width=0.9\linewidth]{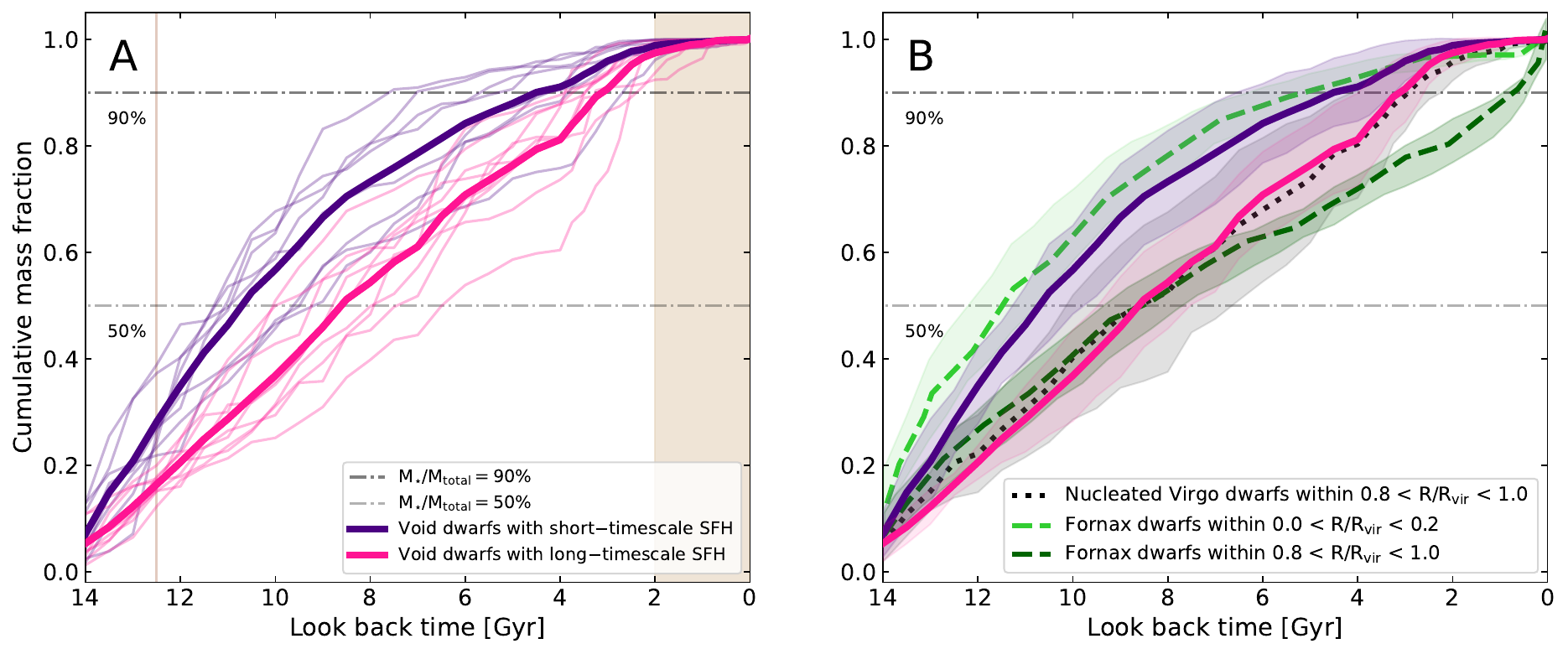}
\caption{The mass assembly history of dwarf galaxies as a function of look back time. Horizontal dashed lines from bottom to top in all panels represent the 50$\%$ and 90$\%$ {cumulative} stellar mass fractions, respectively. \textit{Panel A:} {The brown shaded area highlights the last 2 Gyr.} {The vertical solid line in this panel marks the $\tau_{21.4}$\,$=$\,12.5 Gyr, which is the criteria we adopted to classify galaxies into short- and long-timescale star formation histories, based on \cite{2023Natur.619..269D}}. Eight dwarfs have formed more than 21.4$\%$ of their present-day stellar mass by 12.5 Gyr ago (i.e., $\tau_{21.4}$\,$\geq$\,12.5 Gyr, shown in purple), and nine have $\tau_{21.4}$\,$<$\,12.5 Gyr (shown in pink). Profiles of all dwarfs are shown with narrow lines, colour-coded accordingly, and an average of each sub-sample is shown with the solid thick lines. \textit{Panel B:} Average profiles of isolated and quenched dwarf galaxies (in purple and pink, as in panel A), {and their range of error (shaded area)} are now compared with the average profiles of their counterparts in the core and outskirts of the Fornax cluster \citep[in light and dark green, respectively; taken from][]{2024MNRAS.527.9715R}. Virgo dwarf galaxies hosting an NSC located at large cluster-centric distances (whose average CMFs are shown as dotted black line) exhibit a mass assembly history akin to isolated dwarf galaxies with $\tau_{21.4}$\,$<$\,12.5 Gyr.}
\label{Figure1}
\end{figure*}

{While it is true that quiescent galaxies tend to exhibit redder photometric $\rm g-r$ colours on average, edge-on star-forming galaxies can also appear red due to dust attenuation, as noted by \cite{2012Geha}. To address this concern, we discarded galaxies with H$\rm \alpha$ equivalent width EW(H$\rm \alpha$) $>$ 3 \AA, following the classification by \cite{2011CidFernandes}. This criterion effectively excludes not only star-forming galaxies of various morphologies but also active galaxies hosting BHs.} The EW(H$\rm \alpha$) values used here are taken from the MPA-JHU catalogue of SDSS-DR7 spectra, computed from stellar continuum-subtracted emission lines \citep{2004Tremonti,2004Brinchmann}. This condition implies negligible star formation activity over the past few Myr \citep{2003Bruzual, 2016A&A...590A..44G} and effectively removes AGN hosts \citep{2011CidFernandes}. After applying this condition, we are left with seventeen candidates in voids and five in filaments.

To ascertain the quenched state of star formation activity in these 22 candidates, we analysed their star formation histories (SFHs) using their SDSS-DR7 optical spectra. This step also ensures the removal of nearly edge-on star-forming cases, as discussed in \cite{2012Geha}. Data used in this work were obtained by the 2.5 m telescope at the Apache Point Observatory (APO). SDSS spectra exhibit a wavelength-dependent spectral resolution, with a resolving power of 1500 at $\lambda$ = 3800 $\rm \AA$.  Each spectrum represents the integrated light from the central regions of galaxies, captured within a fibre with an aperture of 3 arcsecs in diameter, encompassing a spatial range of 0.3 to 1.6 kpc (within the 0.01 $<$ z $<$ 0.05). 

\section{Analysis}\label{analysis}
To investigate the SFHs and quantify the quenching timescale of the candidates, we utilized the publicly available full-spectrum fitting routine pyPipe3D \citep{2016RMxAA..52...21S,2016RMxAA..52..171S, 2022Lacerda}. This pipeline linearly combines a set of single stellar population (SSP) models of different ages, metallicities and $\alpha$-enrichment levels (hereafter [$\alpha$/Fe]) to construct the best-fitting synthetic spectrum compared to the observed one (see Fig.~\ref{FigureA2} for an example). Most of the dwarf galaxies investigated here lack prominent emission lines, and the pyPipe3D approach of simultaneously fitting stellar continuum and gas emission lines, removed possible residuals from the stellar continuum, resulting in more accurate estimates of SFHs. 

We utilized \cite{2010Vazdekis,2015Vazdekis} {1272 SSP models} with medium spectral resolution (FWHM = 2.51 \AA) which are based on the MILES stellar library \citep{2006MNRAS.371..703S,2007MNRAS.374..664C,2011A&A...532A..95F}. Here, the SSP models were constructed using a bi-modal initial mass function (IMF) with a slope of 1.3 \citep{1996Vazdekis} and BASTI isochrones \citep{2004Pietrinferni}. The SSP models utilized here span the age range of 0.03 to 14 Gyr, the metallicity range of -2.27$<$[M/H[dex]$<$+0.40, and the $\alpha$-enrichment range of 0.0$<$[$\alpha$/Fe][dex]$<$0.4. We perform the full spectral fitting over the spectral range of 400 to 600 nm, as it contains most of the age and [M/H]-sensitive absorption features in the SDSS wavelength range. Prior to the fitting, we corrected each dwarf's SDSS spectrum for Galactic foreground extinction using the \cite{1989Cardelli} Galactic extinction law, assuming R$_{v}$ = 3.1 and A$_{v}$ values obtained based on galaxy’s right ascension and declination from NASA/IPAC infrared science archive\footnote{\url{https://irsa.ipac.caltech.edu/frontpage/}}. For details of the fitting procedure, see Appendix~\ref{fitquality}.

After each fit, the pipeline returns a list of light-weighted SSP coefficients, representing their contribution to the final synthetic best-fitting spectrum. {These coefficients are the average of multiple fitting iterations the code does internally by randomly varying each pixel of the spectrum based on the errors. The final solution averages all the solutions obtained through these iterations. There is no regularization parameter to fine-tune in pyPipe3D, and the code output is smoothed through these iterations \citep[for more details see][]{2022Lacerda}.} The light-weighted coefficients can then be converted to mass-weighted ones using the mass-to-light ratio (M/L) of each SSP model. By summing the mass coefficient of SSP models with similar ages but different [M/H] and [$\alpha$/Fe] for each galaxy, we constructed its stellar mass fraction as a function of the \textit{look back time} (see Appendix \ref{fitquality}). From this profile, we constructed the cumulative mass fraction (CMF) profiles of the galaxies (in Fig.~\ref{Figure1}). To account for fitting uncertainties, in addition to the internal fitting iterations performed by the code, we conducted 50 independent Monte Carlo (MC) iterations for each galaxy's spectrum. The CMFs reported in this work for each galaxy represent the average of these 50 MC iterations. 

Following the established definition for the quenched population of galaxies \citep[e.g.,][]{2018MNRAS.479.4891F, 2023MNRAS.526.4735F}, we only considered those isolated dwarf galaxies that have formed more than 90 percent of their present-day stellar mass more than 2~Gyr ago (i.e., $\tau_{\rm 90}$ $>$ 2 Gyr) as quenched. By imposing these conditions, we also mitigate the likelihood of these galaxies experiencing episodic star formation since the quenching phase associated with episodic star formation of dwarf galaxies is on the order of Myr \citep[e.g.,][]{2007Stinson}. Additionally, we discarded galaxies with SDSS spectral signal-to-noise ratio (S/N)~$<$~12, for which careful inspection of their fit results revealed unreliable measurements. These two conditions resulted in the removal of five candidates. The final sample consists of 14 galaxies in voids and three galaxies in filaments and their CMFs are shown in Panel A of Fig.~\ref{Figure1}.

\section{Results}\label{results}
\subsection{Analogous quenching in extremely different environments}
The broad range of quenching times shown in panel A of Fig.~\ref{Figure1} (2.2$<$$\tau_{\rm 90}$[Gyr]$<$7.9) suggests that these isolated dwarf galaxies might have undergone various quenching pathways leading to different quenching timescales. {To better assess this hypothesis and guide the discussion, we split these galaxies into two sub-samples based on the cumulative stellar mass they formed by 12.5 Gyr, following \cite{2023Natur.619..269D} classification approach (see Appendix \ref{classification} for details). 
In Fig.~\ref{Figure1}, eight galaxies that formed over 21.4\% of their present-day stellar mass by 12.5 Gyr ago (i.e., $\tau_{\rm 21.4}$ $\geq$ 12.5 Gyr) are classified as ST$-$SFH galaxies, indicated in purple, while the remaining nine are classified as LT$-$SFH galaxies, shown in pink. 

In panel B of Fig.~\ref{Figure1}, we compare the CMFs of isolated and quenched dwarf galaxies with their counterparts in the Fornax cluster. This comparison sample contains 31 early-type dwarf galaxies with log (M$_{\star}$/M$_{\odot}$) $<$ 9.5, located inside Fornax virial radius. These galaxies are selected and investigated by \cite{2024MNRAS.527.9715R}. The CMFs discussed in Fig.~\ref{Figure1} are the average profiles directly adapted from figure 8 of that study and similar to this work, they are derived through the full spectral fitting approach using MILES SSP models. 

On average, the stellar mass assembly history of the discovered isolated and quenched dwarf galaxies (particularly ST$-$SFH ones) resembles that of their counterparts in extremely high-density regions of the Universe (e.g., the core of Fornax cluster as shown in Fig.~\ref{Figure1}, panel B). This is surprising, as cosmic voids and the central regions of massive galaxy clusters differ significantly in numerous aspects. The rapid quenching of star formation in the central regions of clusters is often attributed to strong environmental mechanisms \citep{2022Boselli}. The higher density of the intergalactic medium and the large galaxy population in these regions drive such mechanisms, which are otherwise absent in pristine void environments. This makes environmental effects an unlikely culprit for the similarity in quenching time and overall CMF between the present sample and cluster dwarfs. Even dwarf galaxies located in the cluster outskirts {(i.e., between 0.8 to 1.0 of cluster virial radius, R$_{\rm vir}$)}, where they have inevitably undergone some degree of environmental processing \citep[][]{2019Rhee}, exhibit more recent star formation activity than these counterparts in voids (Panel B of Fig.~\ref{Figure1}).

\begin{figure}
\centering
\includegraphics[scale = 0.44]{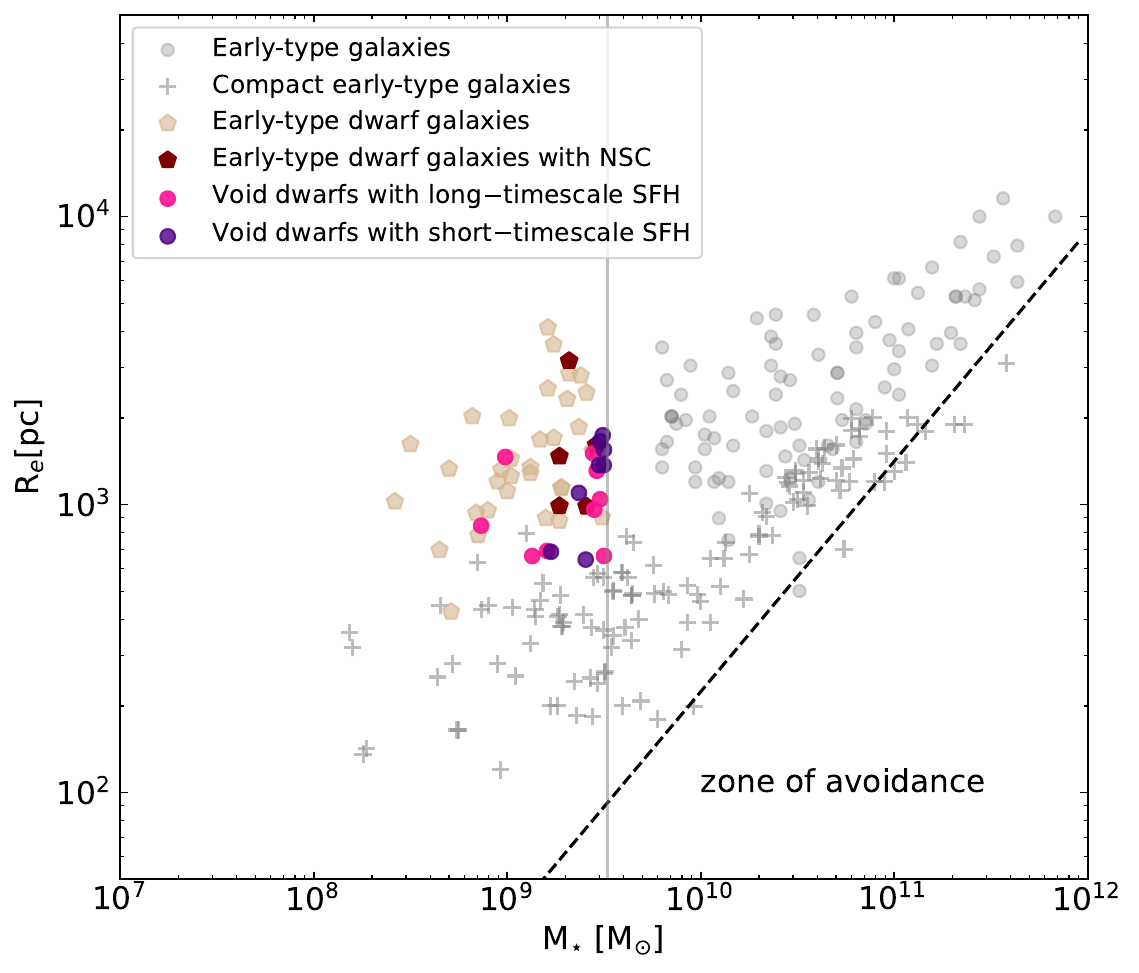}
\caption{The mass-size relation of early-type galaxies. Massive early-type galaxies \citep[shown in gray circles,][]{2016Janz}, compact early-type galaxies \citep[shown in gray plus symbols,][]{2021MNRAS.503.5455F}, and dwarf early-type galaxies in the Virgo cluster \citep[shown in tan coloured symbols,][]{2006Lisker} are compared with the sample of isolated and quenched dwarfs (shown as symbols colour-coded similar to Fig.~\ref{Figure1}). Virgo dwarfs hosting NSC located in the cluster's outskirts, as discussed in Fig.~\ref{Figure1}, are marked in dark red. The stellar masses are taken from the corresponding cited references. {The vertical gray line marks the log(M$_{\star}$/M$_{\odot}$) = 9.5.}  }
\label{Figure2}
\end{figure}

The overall morphologies of detected isolated and quenched dwarf galaxies (see appendix \ref{photometry}) and analysis of their sizes (i.e., in terms of the effective radius) presented in Fig.~\ref{Figure2} indicate that they are comparable to the population of early-type dwarf galaxies found in galaxy groups and clusters. In Fig.~\ref{Figure2}, we compare the size and stellar mass of the main sample with early-type galaxies in the Virgo cluster and compact early-type galaxies. The stellar mass of all Virgo galaxies is taken from the Virgo Cluster Catalogue \citep[VCC;][]{1985Binggeli,2006Lisker}. The effective radius of massive early-type galaxies in Virgo is taken from \cite{2016Janz}. The effective radius of all dwarf galaxies discussed in this figure are measured through the procedure outlined in Appendix \ref{photometry}. The compact early-type sample shown in Fig.~\ref{Figure2} contains 48 galaxies, ranging from massive to dwarf galaxies, compiled from the literature \citep{2016Janz, 2018MNRAS.479.4891F, 2021MNRAS.503.5455F, 2023MNRAS.526.4024G}. The stellar mass and physical size (i.e., effective radius) of these galaxies are taken from the corresponding cited references.  The same plot shows that these galaxies are more extended and inherently distinct than {other isolated quenched systems found to date \citep[e.g. compact ellipticals;][]{2015Chilingarian, 2021MNRAS.503.5455F}}

\subsection{Presence of Nuclear Star clusters}
{Our analysis on the light distribution of the isolated and quenched dwarf galaxies detected here using imaging data and the unsharp masking technique} shows that all of them are Nuclear Star Cluster (NSC) hosts (see Appendix \ref{photometry}). NSCs are very dense and massive stellar assemblies frequently found in the centre of galaxies with M$_{\rm \star} \sim 10^{8-10}\rm\, M_{\odot}$ \citep{2020Neumayer}. {NSCs, with a typical effective radius of 4.4 pc \citep{2020Neumayer}, contribute between 10\% and 50\% of the total light observed within the 3-arcsecond SDSS fibre aperture at the distances of these galaxies \citep[ranging from 48 to 90 Mpc;][]{2012Pan}. It should be noted that given the distance of these galaxies and the typical size of NSCs, what we are observing and discussing (also in Appendix \ref{photometry}) is the broader nuclear regions of these galaxies, which include the NSC along with its immediate surroundings. These nuclei may represent a composite of unresolved central structures and the extended nuclear stellar light. 
Consequently, it remains difficult to draw robust conclusions about the stellar population characteristics of both NSCs and their host galaxies. Given SDSS data coverage, it is also difficult to definitively evaluate the presence or absence of star formation in the unobserved outer regions beyond SDSS fibre coverage}. However, NSCs are generally younger or of similar age compared to their host early-type dwarf galaxies \citep[e.g.,][]{2021Fahrion}. {Through a careful examination of the imaging data for this sample, we found no blue regions indicative of ongoing star formation in either the inner regions or the outskirts of these galaxies. Instead, we observed a smooth, red light distribution with a subtle positive colour gradient extending from the centre to the outskirts. This finding aligns with expectations, as all galaxies in the primary sample belong to the red sequence, where significant star formation is not anticipated. This topic will be explored in greater detail in an upcoming study.}

To investigate possible links between quenching and presence of NSC in these galaxies, we further compare our results with the CMF profiles of NSC hosting early-type dwarf galaxies located at the outskirts of the Virgo cluster in panel B of Fig.~\ref{Figure1}. From 270 NSC hosting galaxies in VCC catalogue, 63 are located within 0.8$<$R/R$_{\rm vir}$$<$1.0. Among these, only five galaxies have stellar masses \citep[taken from][]{2006Lisker} comparable to those investigated in this study and thus are included in the comparison. We computed the average CMF of these galaxies using their SDSS spectra, following the same methodology outlined in Section \ref{analysis}. {Consider that, the CMF of dwarf galaxies located on the outskirts of the Fornax Cluster (dark green dashed profile in panel B of Fig.~\ref{Figure1}), as reported by \cite{2024MNRAS.527.9715R}, represents the average profile of five dwarf galaxies, only one of which hosts a central NSC. Therefore, this profile predominantly reflects the CMF of non-nucleated dwarf galaxies in cluster outskirt.} We found that dwarf galaxies with similar stellar masses, which host a central NSC and are located in the outskirts of the cluster (within 0.8$<$R/R$_{\rm vir}$$<$1.0), exhibit quenching times and mass assembly histories closely resembling those of the isolated and quenched dwarfs in voids and filaments. This suggests a potential connection between quenching mechanisms and the formation of NSCs, which we will discuss in Section \ref{discussion}.

\section{Discussion and conclusion}\label{discussion}
\subsection{What quenched isolated dwarf galaxies in voids and filaments?}

The NSC formation in galaxies with log(M$_{\star}/\rm M_{\odot}) >$ 9.0 could be a consequence of in-situ star formation, where the compression of gas in their gravitational potential well, particularly in its central few parsecs, can trigger episodic and intense star formation bursts and subsequent formation of a central NSC \citep{2020Neumayer, 2022Fahrion}. These intense episodes of star formation can result in strong SNe feedback, stellar winds, and local photoionization, which are all capable of expelling gas and halting further star formation in dwarfs \citep[e.g.,][]{1974Larson}. In particular, \cite{2010Cantalupo} shows that photoionization from local sources, even without strong feedback processes, can regulate the gas accretion and effectively quench star formation in low-mass galaxies. However, our investigations show that not all the dwarf galaxies hosting NSCs in voids and filaments are quenched. 

Mergers can also explain the lack of star formation and the presence of a central NSC in these galaxies \citep{2020Neumayer}. Specifically, zoom-in cosmological simulations of isolated dwarf galaxies reveal that wet mergers between gas-rich dwarfs can significantly increase the central gas density in the merger remnant. This can lead to a pronounced starburst that promotes NSC formation, ultimately quenching subsequent star formation \citep{2024Gray}. However, based on existing theoretical work, NSC formation via self-quenching starburst, triggered by mergers, is more feasible in low-mass dwarf galaxies {(M$_{\star}$$<10^{7}$M$_{\odot}$)} and less likely in galaxies of comparable mass to those discussed in this study \citep{2024Gray}.

Cosmological hydrodynamical simulations propose that Active Galactic Nuclei (AGN)-driven outflows, prompted by BHs in dwarf galaxies, heat up the accreted cold gas and can prevent star formation, independent of their environment \citep[e.g.,][]{2024Arjona-Galvez}. Notably, in dwarf galaxies with stellar masses exceeding 10$^{9}$ M$_{\odot}$ both BHs and NSCs can coexist, yet BHs do not dominate, making their detection more challenging \citep[i.e., log(M$_{\rm BH}$/M$_{\rm NSC}$) $<$ 0,][]{2020Neumayer}. Our selection criteria of EW(H${\alpha}$) $<$ 3 \AA, as discussed in Section \ref{analysis}, decreases the probability of these galaxies being AGN dominated \citep{2011CidFernandes}. Despite that, we could detect AGN through {AGN diagnostic diagrams} only in one galaxy of this sample. Therefore, we cannot overlook the potential influence of BHs in their quenching. This is especially pertinent since the single-fibre diagnostics tools \citep[such as the BPT diagram,][]{1981Baldwin} employed in this work fail to detect off-centre AGN sources, particularly in dwarf galaxies where the signature of their lower luminosity AGN is often obscured by different effects, such as star formation and stellar processes \citep{2015Trump,2024Mezcua}. 

\subsection{The diversity of origins and evolution trajectories}
Similar to \cite{2023Natur.619..269D} results, we also found a clear bimodality in the $\tau_{50}$ and $\tau_{90}$ (i.e., the moment when a galaxy completes the formation of 50\% and 90\% of its current stellar mass, respectively) of the quenched dwarfs (see Panel B of Fig.~\ref{Figure1}). Many studies use $\tau_{50}$ as an observational proxy for the ‘formation time of the galaxy halo’ \cite[e.g.,][]{2017Tojeiro}. Based on this, we can also explore the evolution of each subset separately:

\textbf{ST$-$ SFH:} Galaxies with rapid halo formation (eight isolated dwarfs and all those in the cluster central region) exhibit a consistent pattern of mass assembly throughout time, regardless of their present-day environment. This observation is consistent with the findings of \cite{2023Natur.619..269D}, who showed that, for a much larger sample of galaxies spanning clusters, filaments, and voids, those with ST$-$SFH show no significant differences in their mass assembly. Our results suggest that the processes occurring primarily before  $\tau_{50}$ determine the quenched state of these galaxies today, irrespective of their current environment. The early and rapid episodes of star formation during this phase likely triggered strong SNe feedback and stellar winds, depleting their gas reservoirs, which also leads to NSC formation \citep[][]{2019MNRAS.486L...1S} and fast, early quenching. Dwarf galaxies in central regions of clusters (dashed light green line in Fig.~\ref{Figure1}) have been subjected to harsh environmental conditions for over $\sim$5~Gyr \citep{2019Rhee}. {However, it is less likely that the early quenching observed in isolated void dwarf galaxies reflects pre-processing within a high-density environment, such as a small galaxy group or a tendril \citep[e.g.,][]{2014Alpaslan}. For this to be the case, these galaxies would have had to reside in a dense region that subsequently expanded and dispersed due to the dynamical evolution of the surrounding void \citep[e.g.,][]{2024Kugel}, or alternatively, they would need to have escaped from such high-density regions. Nevertheless, such scenarios, if they occur at all, are exceedingly rare.} If $\tau_{50}$ serves as a proxy for galaxy halo formation, our findings suggest that dwarf galaxies in central regions of clusters formed their halos slightly earlier than their isolated counterparts in voids (solid purple line in Fig.~\ref{Figure1}). This confirms the results of \cite{2004Sheth}, who showed that halos of similar masses tend to form slightly earlier in denser environments.

\textbf{LT$-$SFH:} Galaxies with extended halo formation timescales, including nine isolated dwarfs (solid pink line in Fig.~\ref{Figure1}) discussed here and the average of those in the outskirts of clusters (dark green dashed line, Fig.~\ref{Figure1}), exhibit a consistent mass assembly pattern up to $\tau_{50}$, with no significant differences observed, as in the case of galaxies with ST$-$SFH. However, later in their mass assembly, the CMF of NSC host galaxies in isolation and on the cluster's outskirts begins to diverge from that of non-NSC host galaxies. NSC host dwarfs quench approximately 2~Gyr earlier than their non-NSC host counterparts (dotted black line in Fig.~\ref{Figure1}) in cluster outskirts, underscoring the significant impact of secondary evolutionary processes after $\tau_{50}$ on their mass assembly. These findings suggest that, in some dwarf galaxies, the formation of NSCs could lead to quenching, irrespective of their environment. Alternatively, NSC formation might be a byproduct of other quenching mechanisms, such as AGN activity or intense star formation, which can impact the dwarf galaxy's star formation regardless of its environment. 

\vspace*{5px}
In conclusion, for the first time we have identified a rare population of isolated and quenched dwarf galaxies in cosmic voids and filaments, the existence of which, despite the small sample size, challenges the prevailing understanding of star formation quenching in dwarf galaxies. Each of these dwarfs hosts a central NSC and displays mass assembly and quenching timescales similar to those of dwarf galaxies in clusters. The exact mechanisms responsible for quenching these galaxies remain unclear, as detailed analysis of their physical properties (such as kinematics and stellar population properties) {require additional observational data and sophisticated analysis that are currently not available to us.} Several follow-up studies are planned to explore the proposed quenching mechanisms. These galaxies have the potential to refine cosmological models and the understanding of baryonic physics and are possibly the keys to better constraining the feedback mechanisms that govern the onset and cessation of star formation on small scales.

\begin{acknowledgements}
BB and IP acknowledges financial support from the Grant AST22-4.4, funded by Consejería de Universidad, Investigación e Innovación and Gobierno de España and Unión Europea – NextGenerationEU, and by the research projects PID2020-113689GB-I00 and PID2023-149578NB-I00 financed by MCIN/AEI/10.13039/501100011033. IP acknowledges financial support by the grant FQM108, financed by the Junta de {Andaluc\'ia} (Spain). LSM acknowledges support from Juan de la Cierva fellowship (IJC2019- 041527-I). M.A-F. acknowledges support from ANID FONDECYT iniciaci\'on project
11200107 and the Emergia program (EMERGIA20\_38888) from Consejer\'ia de
Universidad, Investigaci\'on e Innovaci\'on de la Junta de Andaluc\'ia. 
AFM has received support from RYC2021-031099-I and PID2021-123313NA-I00 of MICIN/AEI/10.13039/501100011033/FEDER,UE, NextGenerationEU/PRT. TRL acknowledges support from Juan de la Cierva fellowship (IJC2020-043742-I). 
SDP acknowledges financial support from Juan de la Cierva Formaci\'on fellowship (FJC2021-047523-I) financed by MCIN/AEI/10.13039/501100011033 and by the European Union `NextGenerationEU'/PRTR, Ministerio de Econom\'ia y Competitividad under grants PID2019-107408GB-C44, PID2022-136598NB-C32, and is grateful to the Natural Sciences and Engineering Research Council of Canada, the Fonds de Recherche du Qu\'ebec, and the Canada Foundation for Innovation for funding.
JF-B acknowledges support from the PID2022-140869NB-I00 grant from the Spanish Ministry of Science and Innovation. 
\end{acknowledgements}

\bibliographystyle{aa} 
\bibliography{lettersizes} 

\begin{appendix}
\section{Spectral fitting quality}\label{fitquality}
The SDSS spectra exhibit a wavelength-dependent spectral resolution. Since pyPipe3D performs spectral fitting using a fixed instrumental resolution, we degrade the observed spectra to match the lowest resolution across each galaxy's wavelength range. To achieve this, we calculate the average resolution within 50 \AA\, windows and determine the differential resolution as the quadratic difference between this average and the lowest global resolution. The SDSS spectra are then convolved segment by segment, applying the calculated differential resolution and normalizing across the full spectral range. The final spectral resolution varies between galaxies, ranging from 1.32$<$$\sigma$[\AA]$<$1.75. Similarly, we convolved the SSP models to a resolution comparable to the observed spectra.

An example of the pyPipe3d fit quality on the SDSS spectrum of one of the isolated and quenched dwarf galaxies discussed in this letter is shown in Fig.~\ref{FigureA2}. In {the upper panel} of Fig.~\ref{FigureA3} an example of the stellar light and mass fraction profiles as a function of look back time is shown and errors are the standard deviation of the multiple measurements. This example demonstrates that most of the galaxy's present-day stellar mass was formed over 3 Gyr ago. The peak in the light-weighted star formation rate, occurring $\sim$ 2 Gyr ago, contributed negligibly to the overall stellar mass growth. {As shown in the lower panel of Fig.~\ref{FigureA3}, we find consistent results regarding the global shape of the CMFs and quenching time when using different stellar population synthesis codes, including STECKMAP \citep[][]{2006Ocvirk, 2011MNRAS.415..709S}.}

\begin{figure*}
\centering
\includegraphics[width = 1\textwidth]{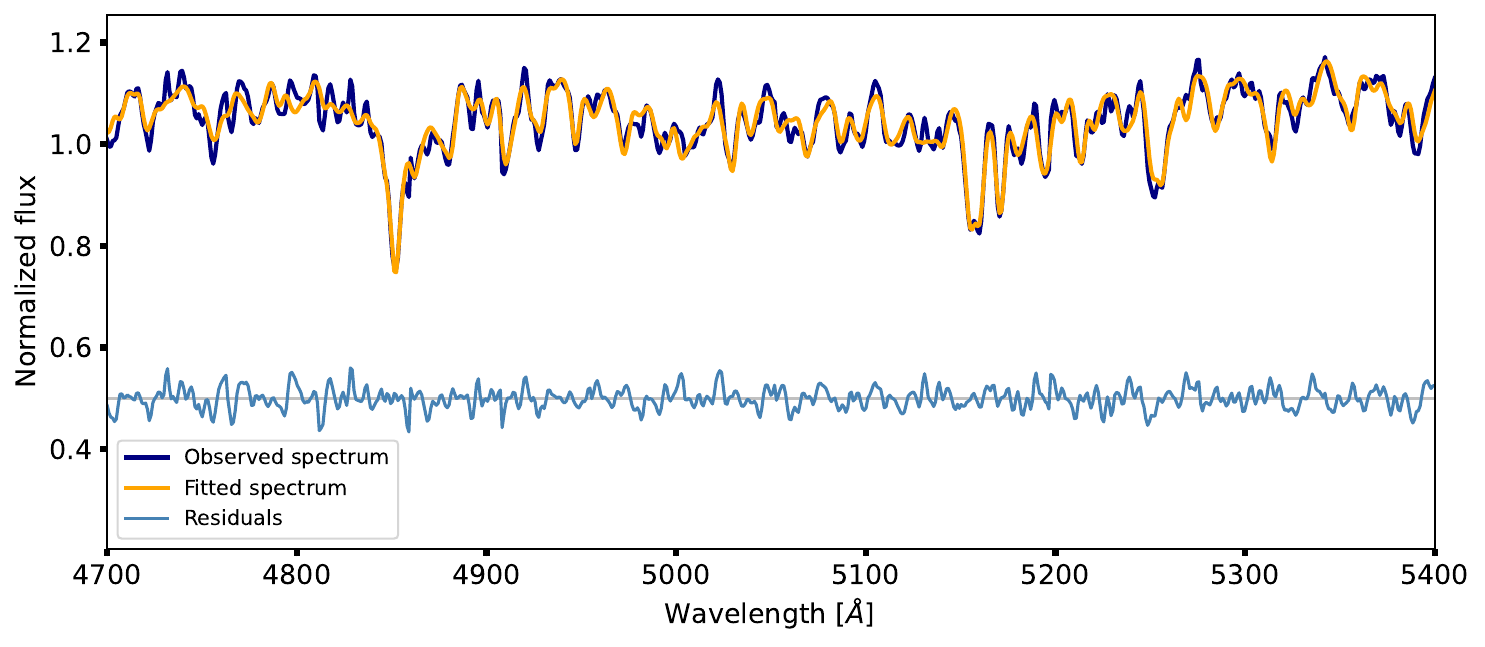}
\caption{The observed spectrum plotted in dark blue is corrected for emission line residuals and is normalized. The pyPipe3d best-fitting synthetic spectrum is shown in orange, and the residuals of the fit are shown in light blue. For better legibility, the residuals are shifted up by 0.5.} 
\label{FigureA2}
\end{figure*}

\begin{figure*}
\centering
\includegraphics[width = 0.8\textwidth]{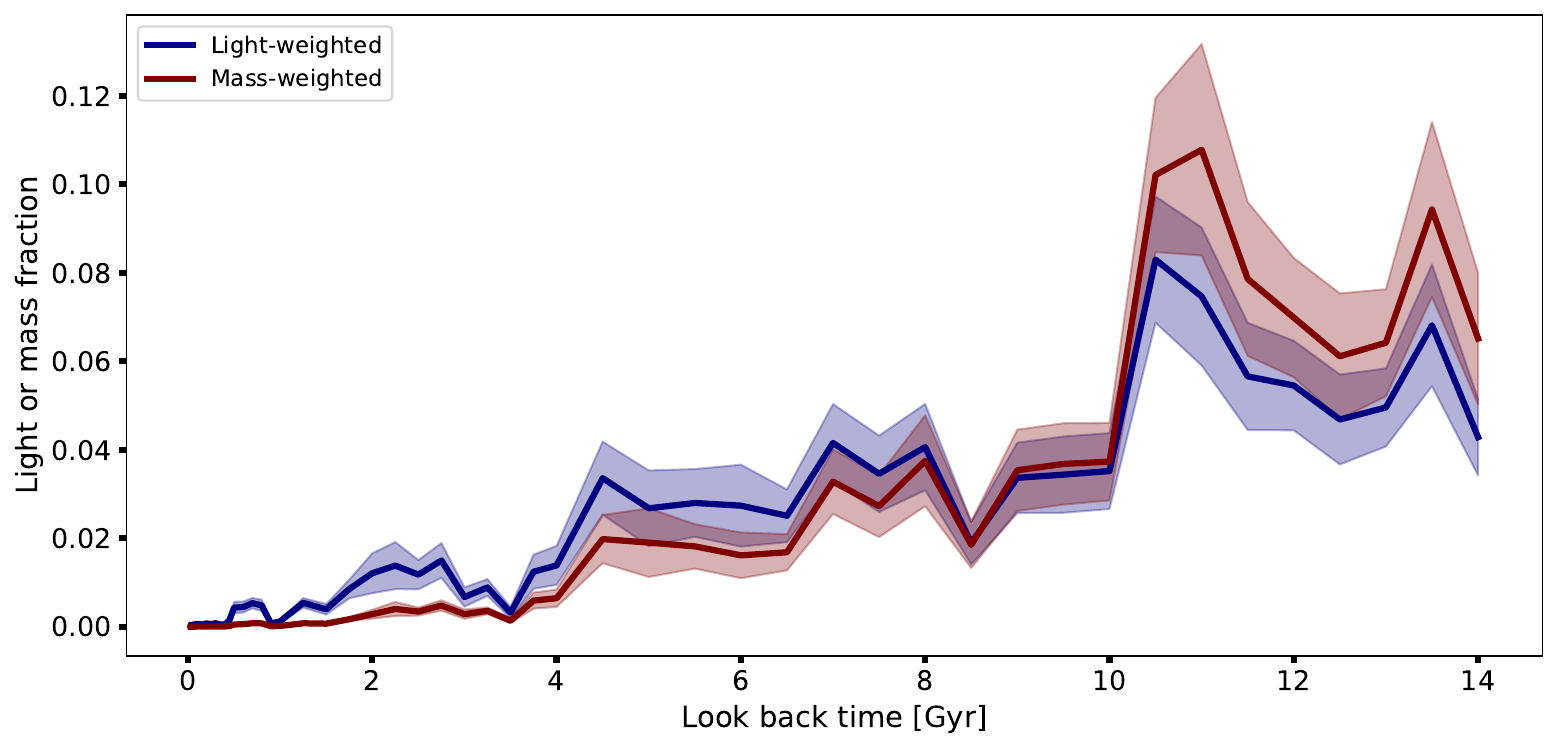}
\includegraphics[width = 0.9\textwidth]{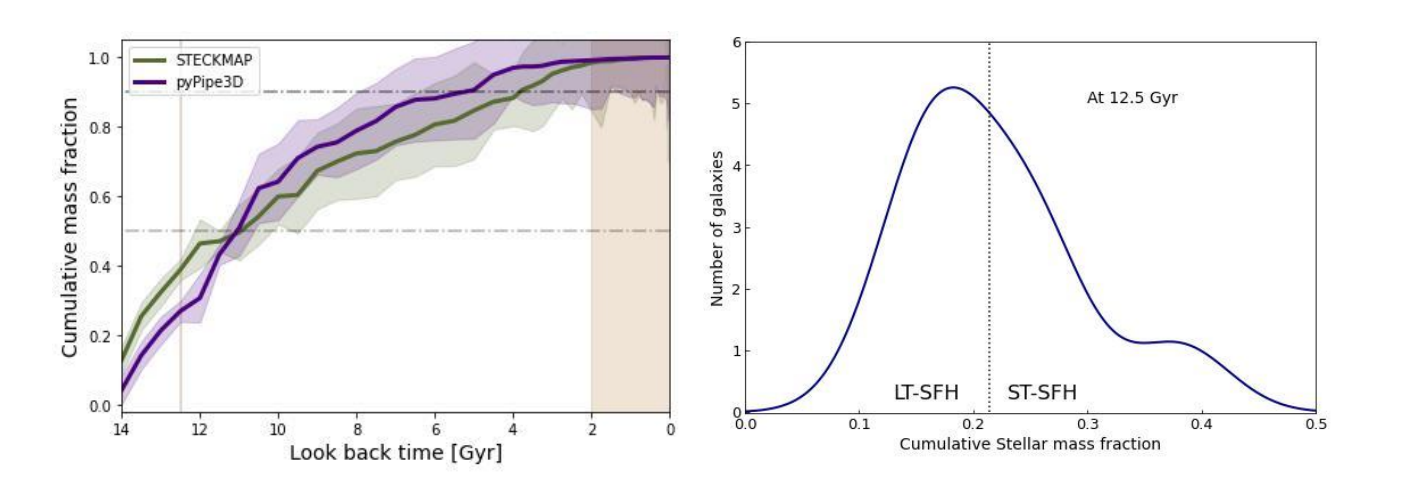}
\caption{The blue and red profiles represent the stellar light and mass fractions formed at different look back times, respectively. \textit{Upper panel:} Each solid line is the average of over 50 profiles constructed from MC iterations using pyPipe3D full spectral fitting routine, and shaded areas around solid lines are the standard deviation of derived profiles. {\textit{Lower left-hand panel:} An example comparison of the CMFs derived for a single galaxy using the pyPipe3D and STECKMAP full spectral fitting routines is presented in purple and green, respectively.} \textit{Lower right-hand panel:} The smoothed distribution of the cumulative star formation histories at 12.5 Gyr. The vertical dashed line marks the 21.4 percent of total stellar mass.}
\label{FigureA3}
\end{figure*}

\section{Short- and Long-Star formation histories}\label{classification}

As explained in Section \ref{results}, the main sample of this work is divided into two sub-groups based on the cumulative stellar mass that each galaxy formed by 12.5 Gyr (indicated by the vertical dotted line in the lower right-hand panel of Fig.~\ref{FigureA3}), following the method and threshold proposed by \cite{2023Natur.619..269D}. In their study, SFHs are classified as short-timescale (ST-SFH) or long-timescale (LT-SFH) based on whether galaxies formed more or less than 21.4\% of their total stellar mass by 12.5 Gyr ago, respectively. The 21.4\% threshold corresponds to a point where the distribution of total stellar mass formed in galaxies 12.5 Gyr ago shows bimodality across different large-scale environments (similar to the lower right-hand panel of Fig.~\ref{FigureA3}).

\section{Analysis of imaging data}\label{photometry}
We utilized r-band imaging data from the Dark Energy Camera Legacy Survey \citep[DECaLS;][]{2016Blum} to investigate the physical size and potential substructures of the main sample discussed in this letter.The FWHM of the measured point spread function (PSF) of the DECaLS r-band images is 1.18 arcseconds \citep{2019Dey}. Similar imaging data were also used to derive the effective radius of Virgo early-type dwarf galaxies within a comparable stellar mass range to the main sample, as discussed in Fig.~\ref{Figure2}. 

To calculate the effective radius, we employed the PetroFit tool, a publicly available fitting routine \citep{2022Geda}, which enables precise photometry, Petrosian profiling, segmentation, and Sersic fitting across various photometric datasets. Using PetroFit's functionalities, we first constructed the growth curve for each galaxy's flux and then calculated its Petrosian properties. In Fig.~\ref{Figure2}, we present the half-light radius for each galaxy. We ensured consistency in the size definitions between our work and the studies we compared with in Fig.~\ref{Figure2}.

\begin{figure*}
\centering
\includegraphics[scale=0.45]{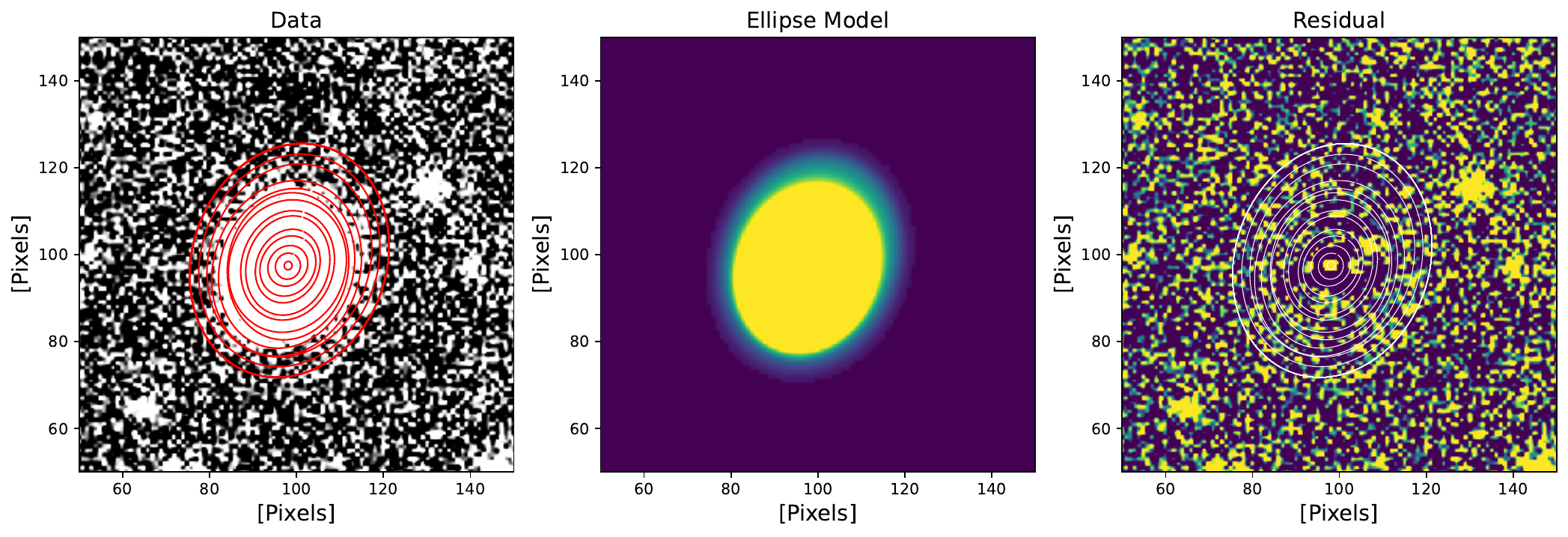}
\includegraphics[scale=0.45]{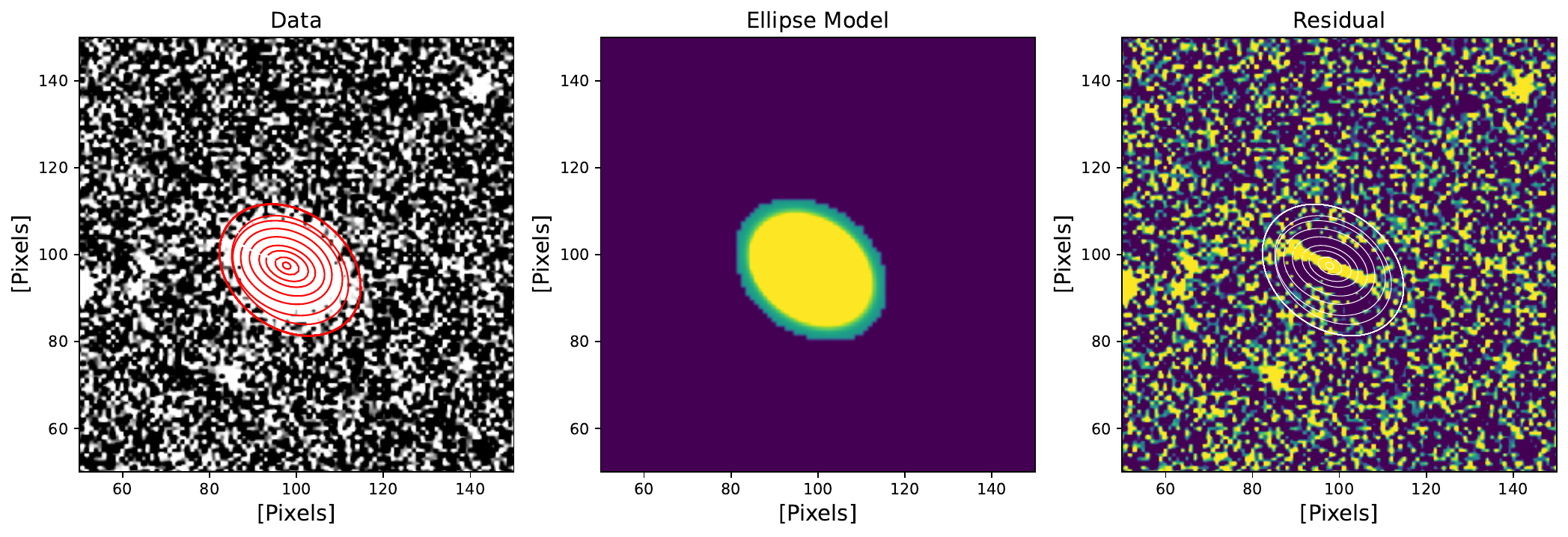}
\caption{Substructures in two isolated and quenched void dwarf galaxies. The left-hand panels show DECaLs r-band data for two examples of the main sample where isophotes of elliptical models are overplotted in red. The middle panel shows the constructed elliptical model for each of the galaxies. The right-hand panels show residual images constructed from subtracting elliptical models from observations. The white contours are isophotes of the elliptical models. The dwarf galaxy shown on the top panels is nucleated, similar to the other 14 galaxies in the main sample. The dwarf in the bottom panels has a disk-like substructure, and three galaxies in the main sample exhibit such substructure. } 
\label{FigureA4}
\end{figure*}

To better investigate various potential substructures hidden within the diffuse light of these galaxies, we employed Photutils \citep{larry_bradley_2024_12585239}, a publicly available photometry routine. With this routine, we applied elliptical isophote fitting on the r-band DECaLs image of each galaxy, generating an elliptical model image that accurately represents its diffuse light. Then, we could detect substructures in these galaxies by subtracting this model from observed data. This simple test shows that all of the galaxies discussed in this study are nucleated (i.e., they host a central NSC), and in three of them, we could also detect a disk-like feature (see Fig.~\ref{FigureA4}). {It should be noted that at the distance of our objects,  NSCs correspond to angular sizes between 0.012  and 0.092 arcseconds, which are significantly smaller than the PSF FWHM. Consequently, the DECaLS images cannot resolve individual NSCs within these galaxies. Instead, we are observing and discussing the broader nuclear regions of these galaxies, which include the NSC along with its immediate surroundings. }

\end{appendix}
\end{document}